\documentclass[12pt]{iopart}

\usepackage{graphicx}
\usepackage{braket}
\usepackage{cite}
\usepackage{dsfont}
\usepackage{amssymb}
\usepackage{mathabx}
\begin{document}

\title{Versatile laser-free trapped-ion entangling gates}

\author{R. T. Sutherland$^{1}$, R. Srinivas$^{2,3}$, S. C. Burd$^{2,3}$, D. Leibfried$^2$, \\ A. C. Wilson$^2$, D. J. Wineland$^{2,3,4}$, D. T. C. Allcock$^{2,3,4}$, \\D. H. Slichter$^2$, S. B. Libby$^1$}

\address{$^1$ Physics Division, Physical and Life Sciences, Lawrence Livermore National Laboratory, Livermore, CA 94550}
\address{$^2$ Time and Frequency Division, National Institute of Standards and Technology, Boulder, CO 80305}
\address{$^3$ Department of Physics, University of Colorado, Boulder, CO 80309}
\address{$^4$ Department of Physics, University of Oregon, Eugene, OR 97403}

\ead{sutherland11@llnl.gov}

\begin{abstract}
We present a general theory for laser-free entangling gates with trapped-ion hyperfine qubits, using either static or oscillating magnetic-field gradients combined with a pair of uniform microwave fields symmetrically detuned about the qubit frequency.  By transforming into a `bichromatic' interaction picture, we show that either ${\hat{\sigma}_{\phi}\otimes\hat{\sigma}_{\phi}}$ or ${\hat{\sigma}_{z}\otimes\hat{\sigma}_{z}}$ geometric phase gates can be performed. The gate basis is determined by selecting the microwave detuning. The driving parameters can be tuned to provide intrinsic dynamical decoupling from qubit frequency fluctuations. The ${\hat{\sigma}_{z}\otimes\hat{\sigma}_{z}}$ gates can be implemented in a novel manner which eases experimental constraints.  We present numerical simulations of gate fidelities assuming realistic parameters.  
\end{abstract}

\maketitle
\section{Introduction}

Due to their inherent uniformity and exceptional coherence properties, trapped ions are a promising platform for scalable quantum simulations and general purpose quantum computing \cite{cirac_1995, monroe_1995,nielsen_2010, haffner_2008,blatt_2008,ladd_2010}. Quantum entanglement, a necessary component of these two applications, is created in the ions' internal degrees of freedom via coupling to shared motional modes \cite{cirac_1995}. This spin-motion coupling is achieved with one or more spatially dependent electromagnetic fields.  One of the critical challenges for trapped-ion quantum logic is obtaining robust, scalable methods for spin-motion coupling with minimal error. The best entangling gate operations to date (fidelity $\approx0.999$) have been implemented using lasers and hyperfine qubits \cite{ballance_2016,gaebler_2016}. In this scheme, two interfering non-copropagating laser beams create a moving optical lattice, whose state-dependent force couples the ions' internal degrees of freedom to their shared motion. The dominant errors reported in Refs. \cite{ballance_2016, gaebler_2016} are due to photon scattering \cite{ozeri_2007}.  An alternative scheme uses microwaves and magnetic field gradients (static or oscillating) to create the desired spin-motion coupling \cite{mintert_2001, ospelkaus_2008, chiaverini_2008, ospelkaus_2011,  khromova_2012, lake_2015, harty_2016, hensinger_2016, wunderlich_2017}. Such laser-free gates are not limited by photon scattering, and phase control is significantly easier than in the optical domain.  Furthermore, microwave and rf sources are readily scalable to meet the requirements of larger quantum processors.

Recently, a microwave-based M\o lmer-S\o renson (${\hat{\sigma}_{\phi}\otimes\hat{\sigma}_{\phi}}$, where $\hat{\sigma}_{\phi}\equiv \hat{\sigma}_{x}\cos\phi + \hat{\sigma}_{y}\sin\phi$) entangling gate \cite{molmer_1999,molmer_2000,roos_2008} was demonstrated with a fidelity of approximately $0.997$~\cite{harty_2016}. This high-fidelity microwave gate, which relies on magnetic field gradients oscillating close to the qubit frequency, was achieved using an additional dynamical decoupling field \cite{viola_1998_pra,viola_1999,bermudez_2012,tan_2013} to suppress errors due to qubit frequency fluctuations, the main source of decoherence in the system. However, the dynamical decoupling demonstrated in Ref.~\cite{harty_2016} requires an extra field that is separate from, and commutes with, the gate Hamiltonian. This increases the experimental complexity as precision phase control of an additional field is required.

Proposals for laser-free ${\hat{\sigma}_{z}\otimes\hat{\sigma}_{z}}$ geometric phase gates~\cite{leibfried_2007,ospelkaus_2008} require an oscillating magnetic field gradient close to the ions' motional frequency.  These gates are appealing because static qubit frequency shifts commute with the gate and can be canceled with a spin-echo sequence\cite{milburn_2000, leibfried_2003}.  However, experimental techniques for generating the necessary gradients usually also result in residual near-resonant electric fields which excite the ion motion and impact gate fidelity \cite{ospelkaus_2008}. These technical challenges limit the implementation of high-fidelity laser-free ${\hat{\sigma}_{z}\otimes\hat{\sigma}_{z}}$ gates.  

Previous laser-free trapped-ion quantum logic experiments with oscillating gradients used a pair of near-field microwave gradients, symmetrically detuned about the qubit frequency, to generate the spin-motion coupling required for an entangling gate \cite{ospelkaus_2011, harty_2016}. To reduce off-resonant qubit excitations and ac Zeeman shifts, the microwave magnetic field was minimized at the position of the ions. Recent theoretical work, however, has shown that gates can still be performed in the presence of microwave fields when the qubits are in the dressed state basis with respect to a \textit{monochromatic} field \cite{wunderlich_2017}. In typical implementations of geometric phase gates, the microwave field is \textit{bichromatic}, which complicates analyzing the gate in the dressed-state basis. 
 
In the work reported here, we derive two-qubit gate dynamics in the interaction picture with respect to the bichromatic microwave field already present in current experimental implementations of geometric phase gates.  We refer to this as the \textit{bichromatic interaction picture}.  We find that the dynamics in this interaction picture produce the same final state as in the laboratory frame, as long as the bichromatic fields are turned on and off adiabatically. For some configurations, the gate basis can be chosen to be either ${\hat{\sigma}_{z}\otimes \hat{\sigma}_{z}}$ or ${\hat{\sigma}_{\phi}\otimes \hat{\sigma}_{\phi}}$ simply by changing the detuning of the bichromatic field. By analyzing these gates in the bichromatic interaction picture, we show that it is possible to dynamically decouple from qubit frequency shifts without adding an extra field.  Finally, we show that this technique enables ${\hat{\sigma}_{z}\otimes\hat{\sigma}_{z}}$ gates with all fields far detuned from the ions' motional frequencies.

The outline of our paper is as follows.  In Sec.~\ref{sec:gates}, we give the theoretical framework for analyzing microwave gates in the bichromatic interaction picture. We then demonstrate how ${\hat{\sigma}_{\phi}\otimes\hat{\sigma}_{\phi}}$ gates, ${\hat{\sigma}_{z}\otimes\hat{\sigma}_{z}}$ gates, and intrinsic dynamical decoupling can be implemented using this framework.  In Sec.~\ref{sec:experimental}, we apply our theory to three experimental situations: a static magnetic field gradient, one that is oscillating close to the qubit frequency, and one that is oscillating close to the motional frequency.  In Sec.~\ref{sec:numerical}, we present numerical calculations of gate fidelities for the near-motional gradient scheme, showing the impact of experimental imperfections on gate performance.  In Sec.~\ref{sec:conclusions}, we present conclusions and prospects for future work.  

\section{Gates in the Bichromatic Interaction Picture}\label{sec:gates}
\subsection{Interaction Picture Dynamics}\label{sec:int_pic_gate}
We assume a Hamiltonian $\hat{H}(t)$, acting on the state $\ket{\psi(t)}$, consisting of two parts:
\begin{eqnarray}
\hat{H}(t) &=& \hat{H}_{\mu}(t) + \hat{H}_{g}(t),
\end{eqnarray}
where we will go into the interaction picture with respect to $\hat{H}_{\mu}(t)$, and $\hat{H}_{g}(t)$ is the remainder of the Hamiltonian. We assume that $\hat{H}_{\mu}(t)$ commutes with itself at all times, and make no such assumption about $\hat{H}_{g}(t)$. Transforming into the interaction picture with respect to $\hat{H}_{\mu}$ gives an interaction picture Hamiltonian $\hat{H}_{I}(t)$:
\begin{eqnarray}\label{eq:interaction_pic}
\hat{H}_{I}(t) &=& \hat{U}^{\dagger}(t)\hat{H}(t)\hat{U}(t) + i\hbar\dot{\;\hat{U}^{\dagger}}(t)\hat{U}(t) \nonumber \\
&=& \hat{U}^{\dagger}(t)\hat{H}_{g}(t)\hat{U}(t),
\end{eqnarray}
where
\begin{eqnarray}\label{eq:unitary_init}
\hat{U}(t) &=& \exp\Big\{-\frac{i}{\hbar}\int^{t}_{0}dt^{\prime}\hat{H}_{\mu}(t^{\prime}) \Big\}.
\end{eqnarray}
In this frame, the time evolution of the transformed state 
\begin{eqnarray}
\ket{\phi(t)} \equiv \hat{U}^{\dagger}(t)\ket{\psi(t)},
\end{eqnarray}
is governed by the interaction picture Schr\"{o}dinger equation
\begin{eqnarray}\label{eq:interact_se}
i\hbar \ket{\dot{\phi}(t)} = \hat{H}_{I}(t)\ket{\phi(t)}.
\end{eqnarray}
After applying $\hat{H}_{I}(t)$ to $\ket{\psi(0)}$ for a duration $t_{f}$, the evolution of $\ket{\psi(t_{f})}$ is described by the unitary propagator $\hat{T}_{I}(t_{f})$ obtained by solving Eq.~(\ref{eq:interact_se}). Thus, the final state in the original frame is given by:
\begin{eqnarray}
\ket{\psi(t_{f})} &= \nonumber & \hat{U}(t_{f})\ket{\phi(t_{f})} \\
&= \nonumber & \hat{U}(t_{f})\hat{T}_{I}(t_{f})\ket{\phi(0)}\\
&= & \hat{U}(t_{f})\hat{T}_{I}(t_{f})\hat{U}^{\dagger}(0)\ket{\psi(0)}.
\end{eqnarray} 
If after the time evolution $\hat{U}(t_{f}) \rightarrow \hat{I}$ ($\hat{U}^{\dagger}(0) = \hat{I}$ trivially), where $\hat{I}$ is the identity operator, we find that
\begin{eqnarray}\label{eq:adiabatic}
\ket{\psi(t_{f})} \rightarrow \hat{T}_{I}(t_{f})\ket{\psi(0)},
\end{eqnarray}
meaning that the propagator in the interaction picture is equal to the propagator in the original frame. 
This well-known result is used extensively in this paper. In Sec.~\ref{sec:shaping} we show that for this system, the desired limit $\hat{U}(t_{f}) \rightarrow \hat{I}$ can be realized by turning $\hat{H}_{\mu}$ on and off adiabatically with pulse shaping.

\subsection{Microwave-Driven Bichromatic Gates}

In this work, we consider a general Hamiltonian for microwave-based gates between $n$ trapped ions with identical frequencies\footnote[1]{We note that this Hamiltonian also describes laser-based gates, but here we only consider microwave-based gates.}:  
\begin{eqnarray}\label{eq:labframe}
\hat{H}_\mathrm{lab}(t) &=& \frac{\hbar \omega_{0}}{2}\hat{S}_{z} + \hbar\omega_r\hat{a}^\dagger\hat{a}+ 2\hbar\Omega_{\mu}\hat{S}_{i}\Big\{\cos([\omega_{0} + \delta]t) + \cos([\omega_{0} - \delta]t) \Big\} \nonumber
\\ && + 2\hbar\Omega_{g}f(t)\hat{S}_{j}\Big\{\hat{a} + \hat{a}^{\dagger} \Big\}.
\end{eqnarray}
\noindent We define $n$-ion Pauli spin operators $\hat{S}_{i} \equiv \sum_{n}\hat{\sigma}_{i,n}$, with $i\in\{x,y\}$ and $j\in\{x,y,z\}$, where $z$ refers to the qubit quantization axis and $\omega_0$ is the qubit frequency. We consider an ion crystal whose internal states are coupled via a motional mode with frequency $\omega_{r}$ and creation (annihilation) operators $\hat{a}^{\dagger}\,(\hat{a})$.  We assume that all other motional modes are sufficiently detuned from $\omega_r$ that they will not couple to the spins.  Here, $\Omega_{\mu,g}$ are Rabi frequencies. The $\Omega_{\mu}$ term represents \textit{two} fields of equal amplitude, detuned from the qubit frequency by $\pm\,\delta$, which only affect the internal states. The $\Omega_{g}$ term couples the internal states and the motion and is implemented with a gradient (along the motional mode) of the $j$ component of a magnetic field.  The time dependence $f(t)$ of the gradient can be an arbitrary function of time; here, we take $f(t)$ to be either constant or sinusoidally oscillating.

We transform Eq.~(\ref{eq:labframe}) into the interaction picture with respect to the ``bare'' ion Hamiltonian $\hat{H}_0 = \hbar\omega_0\hat{S}_z/2 + \hbar\omega_r\hat{a}^\dagger\hat{a}$, and make a rotating wave approximation to eliminate terms near $2\omega_0$, yielding:
\begin{eqnarray}\label{eq:main}
\hat{H}(t)&= \nonumber & \hat{H}_{\mu}(t) + \hat{H}_{g}(t) \\
&= & 2\hbar\Omega_{\mu}\hat{S}_{i}\cos(\delta t) + 2\hbar\Omega_{g}f(t)\hat{S}_{j}\Big\{\hat{a}e^{-i\omega_{r}t} + \hat{a}^{\dagger}e^{i\omega_{r}t} \Big\}.
\end{eqnarray}
We refer to this reference frame as the \textit{ion frame}.  Hamiltonians $\hat{H}_{\mu}(t)$ and $\hat{H}_{g}(t)$, which we refer to  as the \textit{microwave field term} and the \textit{gradient term}, respectively, are the transformed third and fourth terms of Eq.~(\ref{eq:labframe}).  In Eqs.~(\ref{eq:labframe}) and (\ref{eq:main}), the operator $\hat{S}_{j}$ in the gradient term also implicitly incorporates information about the motional mode, and is defined here so that it corresponds to a center-of-mass mode\footnote{For two identical ions it could be trivially extended to an out-of-phase mode by setting $\hat{S}_{j} \equiv \hat{\sigma}_{j,1} - \hat{\sigma}_{j,2}$.}.  For simplicity, we assume that all ions are the same, and can be addressed with a single pair of microwave fields. We note that the following formalism can be generalized to the case of multiple qubit frequencies~\textemdash~either for multiple ion species, or for ions of the same species as discussed for example in Ref.~\cite{khromova_2012}~\textemdash~by using multiple pairs of microwave fields.

\subsubsection{Bichromatic Interaction Picture}\label{sec:interaction_trans}
We now examine the ion frame Hamiltonian from Eq.~(\ref{eq:main}) in the bichromatic interaction picture with respect to the microwave field term $\hat{H}_{\mu}(t)$. This reference frame, rotating at a nonuniform rate, has been utilized in the context of laser-driven gates \cite{roos_2008} to accurately quantify the effect of an off-resonant field. Here, we are interested in analyzing gates in the bichromatic interaction picture itself, as motivated in Sec.~\ref{sec:int_pic_gate}. Note that, for simplicity, we take $\Omega_{\mu}$ to be constant (i.e. we neglect pulse shaping) until Sec.~\ref{sec:shaping}.

We move into the interaction picture with respect to the bichromatic field by making the transformation:
\begin{eqnarray}\label{eq:unitary}
\hat{U}(t) &=& \exp\Big\{-\frac{i}{\hbar}\int^{t}_{0}dt^{\prime}H_{\mu}(t^{\prime}) \Big\} \nonumber \\
&=& \exp\Big\{-2i\Omega_{\mu}\hat{S}_{i} \int^{t}_{0}dt^{\prime}\cos(\delta t^{\prime})\Big\} \nonumber \\
&=& \exp\Big\{-iF(t)\hat{S}_{i}\Big\}.
\end{eqnarray}
Here $F(t) \equiv \frac{2\Omega_{\mu}\sin(\delta t)}{\delta}$.
The interaction picture Hamiltonian is then:
\begin{eqnarray}\label{eq:first_full}
\hat{H}_{I}(t) &=& 2\hbar \Omega_{g}f(t)\Big\{\hat{a} e^{-i\omega_{r}t} + \hat{a}^{\dagger}e^{i\omega_{r}t} \Big\}e^{iF(t)\hat{S}_{i}}\hat{S}_{j}e^{-iF(t)\hat{S}_{i}}. 
\end{eqnarray}
Focusing on the Pauli operators in Eq.~(\ref{eq:first_full}):
\begin{eqnarray}\label{eq:pauli}
&&\nonumber e^{iF(t)\hat{S}_{i}}\hat{S}_{j}e^{-iF(t)\hat{S}_{i}} \\
&=& \nonumber \Big\{ \hat{I}\cos(F(t)) + i\hat{S}_{i}\sin(F(t))\Big\}\hat{S}_{j}e^{-iF(t)\hat{S}_{i}} \\
&=& \hat{S}_{j} + i [\hat{S}_{i},\hat{S}_{j}]\sin(F(t))e^{-iF(t)\hat{S}_{i}}.
\end{eqnarray}
Inserting this into Eq.~(\ref{eq:first_full}) gives:
\begin{eqnarray}\label{eq:second_full}
\hat{H}_{I}(t) &=& 2\hbar \Omega_{g}f(t)\Big\{\hat{a} e^{-i\omega_{r}t} + \hat{a}^{\dagger}e^{i\omega_{r}t} \Big\}\Big\{\hat{S}_{j} + i [\hat{S}_{i},\hat{S}_{j}]\sin(F(t))e^{-iF(t)\hat{S}_{i}}\Big\}. \nonumber \\
\end{eqnarray}
If $i = j$, then Eq.~(\ref{eq:pauli}) $\rightarrow \hat{S}_{j}$, and $\hat{H}_{I}(t)$ is equal to $\hat{H}_{g}(t)$. However, if $i\neq j$, then Eq.~(\ref{eq:second_full}) becomes:
\begin{eqnarray}\label{eq:noncom}
\hat{H}_{I}(t) &=& 2\hbar \Omega_{g}f(t)\Big\{\hat{a} e^{-i\omega_{r}t} + \hat{a}^{\dagger}e^{i\omega_{r}t} \Big\}\Big\{ \hat{S}_{j}\cos(2F(t)) - \epsilon_{ijk}\hat{S}_{k}\sin(2F(t)) \Big\}.\nonumber \\
\end{eqnarray}
Using the Jacobi-Anger expansion \cite{abramowitz_1967}, we obtain:
\begin{eqnarray}\label{eq:jacobi_expand}
\hat{H}_{I}(t) &=& 2\hbar \Omega_{g}f(t)\Big\{\hat{a} e^{-i\omega_{r}t} + \hat{a}^{\dagger}e^{i\omega_{r}t} \Big\}\Big\{ \hat{S}_{j}\Big[J_{0}\Big(\frac{4\Omega_{\mu}}{\delta}\Big) + 2\sum_{n=1}^{\infty}J_{2n}\Big( \frac{4\Omega_{\mu}}{\delta}\Big)\cos(2n\delta t) \Big] \nonumber \\ 
&& - 2\epsilon_{ijk}\hat{S}_{k}\sum^{\infty}_{n=1}J_{2n-1}\Big(\frac{4\Omega_{\mu}}{\delta}\Big)\sin([2n-1]\delta t) \Big\}, 
\end{eqnarray}
where $J_{n}$ is the $n^{th}$ Bessel function, and $\epsilon_{ijk}$ is the Levi-Civita symbol. We consider two possible functional forms of $f(t)$: sinusoidal, corresponding to the oscillating magnetic field gradient from an ac-current-carrying wire \cite{ospelkaus_2011,harty_2016}, or constant, due to the magnetic field gradient induced by a permanent magnet \cite{lake_2015, weidt_2016, khromova_2012} or a dc-current-carrying wire \cite{welzel_2018}.

\begin{figure}[h]
\includegraphics[width=0.75\textwidth]{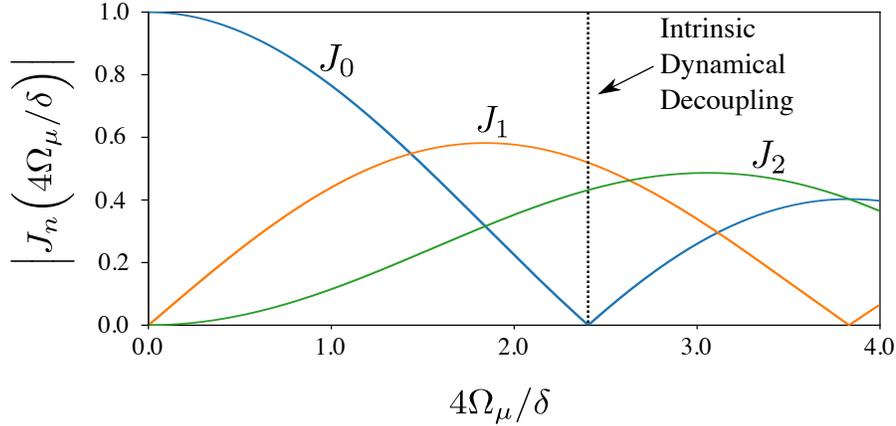}
\centering
\caption{Relative strengths of the gate Rabi frequencies versus $4\Omega_{\mu}/\delta$ for the first three resonances in the bichromatic interaction picture when the microwave field term ($\propto \,\hat{S}_{i}$) does not commute with the gradient term ($\propto \,\hat{S}_{j}$) in the Hamiltonian. Note that at the point where intrinsic dynamical decoupling occurs (dotted line), the values of the $J_{1,2}$ Rabi frequencies are near their maximum values.}
\label{fig:bessels}
\end{figure}

When $i\neq j$, Eq.~(\ref{eq:jacobi_expand}) shows an infinite series of resonances in the bichromatic interaction picture, each with a strength proportional to a Bessel function. We can choose specific values of $\omega_r$, $\delta$, and $n$ with a given $f(t)$ such that one of these terms in Eq.~(\ref{eq:jacobi_expand}) is resonant, i.e. stationary or slowly varying in time. In typical schemes $\delta \gg \Omega_{g}f(t)$~\cite{ospelkaus_2008,ospelkaus_2011, harty_2016}, such that near any particular resonance, one can ignore the off-resonant terms in Eq.~(\ref{eq:jacobi_expand}), which scale as $(\Omega_g f(t)/\delta)^2$\footnote[1]{In the case of multiple qubit frequencies, there will be additional terms in Eq.~(\ref{eq:jacobi_expand}) at other frequencies.  Whether or not these terms can be neglected will depend on the specific values of the qubit frequencies as well as $\delta$, $\Omega_g$, and $f(t)$.}. Further examination reveals that even Bessel function resonances correspond to gate operations where the spin operator $\hat{S}_{j}$ for the gate is identical to the spin operator for the gradient term in Eq.~(\ref{eq:main}). The odd Bessel function resonances correspond to gates whose spin operator $\hat{S}_{k}$ is orthogonal to both the microwave and gradient spin operators $\hat{S}_{i}$ and $\hat{S}_{j}$, respectively. In the typical case, $i \in \{x,y\}$ and $j=z$, this will result in the even and odd Bessel function resonances corresponding to ${\hat{\sigma}_{z}\otimes\hat{\sigma}_{z}}$ and ${\hat{\sigma}_{\phi}\otimes\hat{\sigma}_{\phi}}$ gates (specifically, $\hat{\sigma}_{y}\otimes\hat{\sigma}_{y}$ or $\hat{\sigma}_{x}\otimes\hat{\sigma}_{x}$ gates, depending on the choice of $i$). Figure~\ref{fig:bessels} shows the relative Rabi frequencies of the gates corresponding to the first three resonances versus $4\Omega_{\mu}/\delta $.

\subsubsection{Intrinsic Dynamical Decoupling}\label{sec:dd}
Dynamical decoupling \cite{viola_1998_pra,viola_1999,uhrig_2007} is a useful tool for error suppression in trapped-ion quantum logic experiments \cite{biercuk_2009,timoney_2011,bermudez_2012,piltz_2013,tan_2013,harty_2016,manovitz_2017}. For example, Ref.~\cite{harty_2016} achieved an entangling gate fidelity of approximately $0.997$ by using continuous dynamical decoupling, making the gate operation highly insensitive to qubit frequency fluctuations.  This was done by adding an oscillating magnetic field at the qubit frequency that commutes with the gate but not with qubit frequency fluctuations, thus suppressing the leading source of error while leaving the gate unaffected. Analysis in the bichromatic interaction picture, however, shows it is possible to perform a dynamically decoupled entangling gate operation without adding an extra field, simplifying the experimental setup.

We illustrate this intrinsic dynamical decoupling by adding an error term to the Hamiltonian shown in Eq.~(\ref{eq:main}):
\begin{eqnarray}\label{eq:error_term}
\hat{H}_{z} = \frac{\hbar\varepsilon }{2}\hat{S}_{z},
\end{eqnarray}
where $\varepsilon$ is a (possibly time-dependent) qubit frequency shift, arising for example from environmental noise, control field fluctuations, or miscalibration of the qubit frequency.  Assuming $i \in \{x,y\}$, transforming this term into the bichromatic interaction picture gives:
\begin{eqnarray}\label{eq:dephased_zeeman}
\hat{H}_{I,z} &=& \frac{\hbar\varepsilon}{2}\Big\{ \hat{S}_{z}\Big[J_{0}\Big(\frac{4\Omega_{\mu}}{\delta}\Big)  + 2\sum_{n=1}^{\infty}J_{2n}\Big( \frac{4\Omega_{\mu}}{\delta}\Big)\cos(2n\delta t) \Big]  \nonumber \\ && + 2\epsilon_{ikz}\hat{S}_{k}\sum^{\infty}_{n=1}J_{2n-1}\Big(\frac{4\Omega_{\mu}}{\delta}\Big)\sin([2n-1]\delta t) \Big\}. 
\end{eqnarray}
If $\varepsilon$ varies slowly on timescales of $1/\delta$, then the only term in $\hat{H}_{I,z}$ that is not oscillating near a multiple of $\delta$ is $\propto \, J_{0}( 4\Omega_{\mu}/\delta)$. Therefore, if we set $4\Omega_{\mu}/\delta \simeq 2.405$, the first zero of the $J_{0}$ Bessel function, we leave only fast-oscillating qubit frequency shift terms which contribute negligible dephasing (scaling as $(\varepsilon/\delta)^2$). Fortunately, the value of $4\Omega_{\mu}/\delta$ where dynamical decoupling is achieved occurs near the maxima of the $J_{1}$ and $J_{2}$ Bessel functions, so operating here only results in a modest reduction in gate speed of $\approx11\,\%$ relative to the fastest achievable $J_{1}$ and $J_{2}$ gates.

\subsubsection{Adiabatic Pulse Shaping}\label{sec:shaping}
In this section, we show that if the microwave bichromatic field is smoothly ramped on and off over a time $\tau\gg 2\pi/\delta$, the final wave function in the ion frame approaches the final wave function in the bichromatic interaction picture. In other words, the unitary transformation defined by Eq.~(\ref{eq:unitary_init}) approaches the identity, $\hat{U}(t_{f})\rightarrow \hat{I}$.

To model microwave pulse shaping, we modify the microwave field term in Eq.~(\ref{eq:main}) to include a time-dependent envelope $g(t)$ with a continuous first derivative:
\begin{equation}
\hat{H}_{\mu}(t) \rightarrow 2\Omega_{\mu}g(t)\cos(\delta t)\hat{S}_{i},
\end{equation}
where $g(t)$ is assumed to vary slowly on the timescale $2\pi/\delta$. The following assumptions about the pulse shape are also made:
\begin{eqnarray}\label{eq:pulse}
g(t = 0, t_{f}) &= \nonumber & 0 \\
g(\tau \leq t \leq t_{f} - \tau) &=  & 1,
\end{eqnarray}
where $t_{f}$ is the final gate time. In words, this assumes that the microwave Rabi frequency is equal to zero at the beginning and end of the gate operation, and is constant in between the ramps. At the end of the gate operation, the unitary transformation into the bichromatic interaction picture is:
\begin{eqnarray}\label{eq:unitary_pulse}
\hat{U}(t_{f}) = \exp\Big\{-i\int^{t_{f}}_{0}dt^{\prime} 2\Omega_{\mu}g(t^{\prime})\cos(\delta t^{\prime})\hat{S}_{i}\Big\}.
\end{eqnarray}
Integrating by parts gives:
\begin{eqnarray}\label{eq:unit_integral}
\hat{U}(t_{f}) = \exp\Big\{ \frac{2i\Omega_{\mu}}{\delta}\Big(\int^{\tau}_{0}dt^{\prime}\dot{g}(t^{\prime})\sin(\delta t^{\prime}) + \int^{t_{f}}_{t_{f}-\tau}dt^{\prime}\dot{g}(t^{\prime})\sin(\delta t^{\prime})\Big)\hat{S}_{i}\Big\}.
\end{eqnarray}
If $\dot{g}(t)$ is a slowly varying function with respect to $\sin(\delta t)$, then the larger the value of $\tau$ is relative to $2\pi/\delta$, the smaller the values of the two integrals in Eq.~(\ref{eq:unit_integral}). Thus, in the limit $\tau \gg 2\pi/\delta$, $\hat{U}(t_{f})\rightarrow \hat{I}$, and the final ion frame state approaches the final interaction picture state. Note that this effect is independent of the actual shape of the pulse envelope, provided it is slowly varying. The effect of pulse shaping is discussed for a specific example in Sec.~\ref{sec:phi_gate}, and shown in Fig.~\ref{fig:pulse_ms}. We also point out that pulse shaping will slightly change the optimal gate times due to the changing Rabi frequency during the rise and fall times. 

\section{Experimental Methods of Implementation}\label{sec:experimental}
Depending on the choice of $\hat{S}_{i}$, $\hat{S}_{j}$, and the field gradient function $f(t)$, the preceding derivation can be applied to many experimental systems. In this section, we describe three varieties of microwave-based entangling gates using this formalism.

\subsection{Static Gradient}
A well-studied microwave spin-motion coupling scheme uses a static magnetic field gradient in combination with one or more microwave fields~\cite{mintert_2001}. One previous demonstration of this scheme uses a pair of microwave fields symmetrically detuned about the qubit frequency~\cite{lake_2015}. The ion frame Hamiltonian is then:
\begin{equation}\label{eq:static_rwa}
\hat{H}(t) = 2\hbar\Omega_{\mu}\hat{S}_{x}\cos(\delta t) + 2\hbar\Omega_{g}\hat{S}_{z}\Big\{\hat{a} e^{-i\omega_{r}t} + \hat{a}^{\dagger}e^{i\omega_{r}t} \Big\}.
\end{equation}
This system corresponds to $\hat{S}_{i} = \hat{S}_{x}$, $\hat{S}_{j} = \hat{S}_{z}$, and $f(t) = 1$ in Eq.~(\ref{eq:main}). With these choices, Eq.~(\ref{eq:jacobi_expand}) becomes
\begin{eqnarray}\label{eq:static_jacobi}
\nonumber \\
\hat{H}_{I}(t) &=& 2\hbar \Omega_{g}\Big\{\hat{a} e^{-i\omega_{r}t} + \hat{a}^{\dagger}e^{i\omega_{r}t} \Big\}\Big\{ \hat{S}_{z}\Big[J_{0}\Big(\frac{4\Omega_{\mu}}{\delta}\Big)+ 2\sum_{n=1}^{\infty}J_{2n}\Big( \frac{4\Omega_{\mu}}{\delta}\Big)\cos(2n\delta t) \Big] \nonumber
\\ && + 2\hat{S}_{y}\sum^{\infty}_{n=1}J_{2n-1}\Big(\frac{4\Omega_{\mu}}{\delta}\Big)\sin([2n-1]\delta t) \Big\}.
\end{eqnarray}
If $\Omega_{g} \ll \delta$, we keep only the near resonant terms in this equation. We obtain a ${\hat{\sigma}_{z}\otimes \hat{\sigma}_{z}}$ gate when $2n\delta \sim \omega_{r}$, and a ${\hat{\sigma}_{\phi}\otimes\hat{\sigma}_{\phi}}$ gate (specifically, a ${\hat{\sigma}_{y}\otimes\hat{\sigma}_{y}}$ gate), when $(2n -1)\delta \sim \omega_{r}$.

\subsection{Near-Qubit-Frequency Oscillating Gradient}\label{sec:dynamic_grad}
Another method for spin-motion coupling uses a near-field gradient oscillating close to the qubit frequency \cite{ospelkaus_2008,ospelkaus_2011}. Since the gradient and the microwave term are caused by the same field, we take them to point in the same direction.  The ion frame Hamiltonian is then given by: 
\begin{equation}
\hat{H}(t) = 2\hbar\Omega_{\mu}\cos(\delta t)\hat{S}_{x} + 2\hbar\Omega_{g}\cos(\delta t)\hat{S}_{x}\Big\{ \hat{a} e^{-i\omega_{r}t} + \hat{a}^{\dagger}e^{i\omega_{r}t}\Big\}.
\end{equation}
This system corresponds to $\hat{S}_{i} = \hat{S}_{x}$, $\hat{S}_{j} = \hat{S}_{x}$, and $f(t) = \cos(\delta t)$ in Eq.~(\ref{eq:main}).  Since the microwave term commutes with the gradient term, the bichromatic interaction picture Hamiltonian is simply:
\begin{equation}\label{eq:near_qubit_ip}
\hat{H}_{I}(t) =  2\hbar\Omega_{g}\cos(\delta t)\hat{S}_{x}\Big\{ \hat{a} e^{-i\omega_{r}t} + \hat{a}^{\dagger}e^{i\omega_{r}t}\Big\}.
\end{equation}
This Hamiltonian realizes a ${\hat{\sigma}_{\phi} \otimes \hat{\sigma}_{\phi}}$ gate (specifically, a ${\hat{\sigma}_{x}\otimes\hat{\sigma}_{x}}$ gate); the infinite series of resonances in Eq. (\ref{eq:jacobi_expand}) is absent because the microwave term and the gradient term commute ($i=j$).  In the presence of a qubit frequency shift of the form in Eq. (\ref{eq:error_term}), transforming into the bichromatic interaction picture will then add a term to Eq. (\ref{eq:near_qubit_ip}) of the form shown in Eq. (\ref{eq:dephased_zeeman}); the same analysis from Sec.~\ref{sec:dd} regarding intrinsic dynamical decoupling then applies.

\subsection{Near-Motional-Frequency Oscillating Gradient}\label{sec:near_motion}
Spin-motion coupling can also be accomplished via separate gradient and microwave fields oscillating at near-motional and near-qubit frequencies, respectively. This was demonstrated in Ref.~\cite{ding_2014} by using a running optical lattice to create an oscillating gradient of the ac Stark shift near the ion motional frequencies. Another possibility is to superimpose separate near-qubit and near-motional frequency currents on near-field electrodes in a surface electrode trap \cite{srinivas_2018}. Choosing the gradient to lie along the quantization axis and the microwave fields to be perpendicular to the quantization axis gives the ion frame Hamiltonian
\begin{eqnarray}\label{eq:dynamic_ham}
\hat{H}(t) = 2\hbar\Omega_{\mu}\cos(\delta t)\hat{S}_{x} + 2\hbar\Omega_{g}\cos(\omega_{g} t)\hat{S}_{z}\Big\{ \hat{a}e^{-i\omega_{r}t} + \hat{a}^{\dagger}e^{i\omega_{r}t}\Big\},
\end{eqnarray}
where $\omega_{g}$ is the frequency of the oscillating gradient field. We identify $\hat{S}_{i} = \hat{S}_{x}$, $\hat{S}_{j} = \hat{S}_{z}$, and $f(t) = \cos(\omega_g t)$; using these choices, Eq.~(\ref{eq:jacobi_expand}) becomes:
\begin{eqnarray}\label{eq:dynamic_jacobi}
\hat{H}_{I}(t) &=& 2\hbar \Omega_{g}\cos(\omega_{g}t)\Big\{\hat{a} e^{-i\omega_{r}t} + \hat{a}^{\dagger}e^{i\omega_{r}t} \Big\}\Big\{ \hat{S}_{z}\Big[J_{0}\Big(\frac{4\Omega_{\mu}}{\delta}\Big)  \nonumber \\ 
&&+ 2\sum_{n=1}^{\infty}J_{2n}\Big( \frac{4\Omega_{\mu}}{\delta}\Big)\cos(2n\delta t) \Big] + 2\hat{S}_{y}\sum^{\infty}_{n=1}J_{2n-1}\Big(\frac{4\Omega_{\mu}}{\delta}\Big)\sin([2n-1]\delta t) \Big\}. \nonumber \\
\end{eqnarray}
This is similar to the static field case of Eq.~(\ref{eq:static_jacobi}), only with resonances occurring when $\delta$ is an integer multiple of $|\omega_{r}\pm\omega_{g}|$, rather than $\omega_{r}$. As a result, the Bessel function extrema and roots can be reached with lower $\Omega_{\mu}$ than for the static or near-qubit frequency gradient cases.  For the numerical demonstrations presented in Sec.~\ref{sec:numerical}, we will use this near-motional gradient scheme\textemdash relevant to recent experiments~\cite{srinivas_2018}\textemdash as an example.

\section{Numerical Demonstrations}\label{sec:numerical}

\subsection{${\hat{\sigma}_{\phi}\otimes\hat{\sigma}_{\phi}}$ Gate}\label{sec:phi_gate}
We numerically demonstrate the main results of this work using the system described in Sec.~\ref{sec:near_motion}, where a pair of microwave fields, oscillating near the qubit frequency and polarized in the $\hat{x}$ direction, are combined with a gradient field oscillating near the motional frequency and polarized in the $\hat{z}$ direction. The qualitative results demonstrated below apply to all schemes presented in Sec.~\ref{sec:experimental}, however. If we set $\delta\sim(\omega_{r} - \omega_{g})$, only keeping the resonant terms in Eq.~(\ref{eq:dynamic_jacobi}) gives
\begin{equation}\label{eq:ms_gate_int}
\hat{H}_{I}(t) \simeq i\hbar\Omega_{g}J_{1}\Big(\frac{4\Omega_{\mu}}{\delta}\Big)\hat{S}_{y}\Big\{\hat{a}^{\dagger} e^{-i\Delta t}  - \hat{a}e^{i\Delta t}\Big\},
\end{equation}
where $\Delta \equiv \delta - (\omega_{r} - \omega_{g})$.  Eq. (\ref{eq:ms_gate_int}) corresponds to a ${\hat{\sigma}_{\phi}\otimes\hat{\sigma}_{\phi}}$ gate (specifically, a ${\hat{\sigma}_{y}\otimes\hat{\sigma}_{y}}$ gate) with a Rabi frequency of $\Omega_{\phi} \equiv \Omega_{g}J_{1}\Big( 4\Omega_{\mu}/\delta \Big)$. While the time propagator for the ion frame Hamiltonian (Eq.~(\ref{eq:dynamic_ham})) is fairly complicated to solve analytically, the time propagator for this interaction picture Hamiltonian is well-known \cite{roos_2008,molmer_1999,solano_1999,molmer_2000}. At  $t_{f} = 2\pi /\Delta$ the propagator is
\begin{equation}
\hat{T}_{I}(t_{f}) = \exp\Big\{ -\frac{2\pi i}{\Delta^{2}}(\Omega_{\phi}\hat{S}_{y})^{2}\Big\}.
\end{equation}
For a system starting in the ground state $\ket{\downarrow\downarrow}$, this gate generates a maximally entangled Bell state when $\Delta = 4\Omega_{\phi}$:
\begin{equation}\label{eq:bell}
\ket{Bell}\equiv \frac{1}{\sqrt{2}}\{\ket{\downarrow\downarrow} + i \ket{\uparrow\uparrow}\},
\end{equation}
ignoring an overall phase.
\begin{figure}[h]
\includegraphics[width=0.9\textwidth]{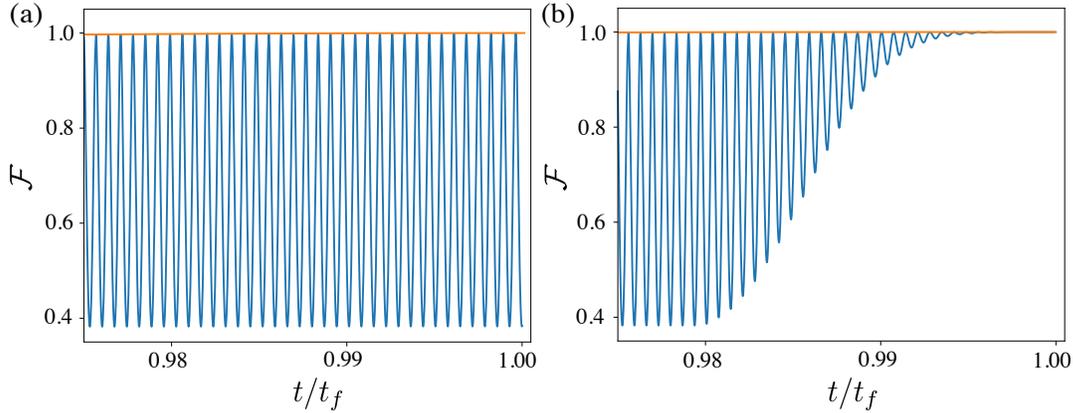}
\centering
\caption{Numerical simulation of the fidelity $\mathcal{F}$ of the maximally entangled Bell state of Eq.~(\ref{eq:bell}) versus time (normalized to $t_{f}$), for the ${\hat{\sigma}_{\phi}\otimes\hat{\sigma}_{\phi}}$ gate described in this section. In both panels, the high frequency blue line corresponds to Eq.~(\ref{eq:dynamic_ham}), i.e. the ion frame Hamiltonian, and the  the orange line corresponds to Eq.~(\ref{eq:ms_gate_int}), i.e. the bichromatic interaction picture Hamiltonian. Panel (a) shows a gate with no pulse shaping, where large-amplitude oscillations at $\delta$ make the ion frame gate fidelity highly sensitive to the exact value of $t_{f}$.  Panel (b) shows the same gate operation including a $\tau = 10\,\mu$s Blackman envelope at the beginning and the end of the gate sequence; the ion frame fidelity smoothly approaches the interaction picture fidelity at the end of the gate.}
\label{fig:pulse_ms}
\end{figure}

The fidelity $\mathcal{F} \equiv \bra{Bell}\hat{\rho}(t)\ket{Bell}$ of this entangling gate is shown in Fig.~\ref{fig:pulse_ms}, where $\hat{\rho}(t)$ is the reduced density operator for the qubit subspace. We simulate this gate operation for a two-ion system undergoing the dynamics caused by the full ion frame Hamiltonian, Eq.~(\ref{eq:dynamic_ham}), as well as the bichromatic interaction picture Hamiltonian, Eq.~(\ref{eq:ms_gate_int}). This is done using realistic experimental parameters of $\Omega_{\mu}/2\pi = 500$ kHz, $\Omega_{g}/2\pi = 1$ kHz, $\omega_{r}/2\pi = 6.5$~MHz and $\omega_{g}/2\pi = 5$~MHz. Figure~\ref{fig:pulse_ms}(a) shows the gate fidelity in both the bichromatic interaction picture and the ion frame, without pulse shaping. In the interaction picture (i.e. $\ket{\phi(t)}$ as opposed to $\ket{\psi(t)}$ from Sec.~\ref{sec:int_pic_gate}), the state $\ket{Bell}$ is created with $\mathcal{F} = 1$. However, in the ion frame, the fidelity is oscillating according to $\mathcal{F} \propto \cos^{4}\big(\frac{2\Omega_{\mu}}{\delta}\sin\{\delta t\}\big)$ (see appendix). Fig.~\ref{fig:pulse_ms}(b) shows that, as described in Sec.~\ref{sec:shaping}, with sufficient pulse shaping the ion frame and rotating frame fidelities converge at the end of the gate. These simulations implement a microwave envelope $g(t)$ with a $10 ~\mu s$ Blackman rising and falling edge \cite{blackman_1958}. Thus, even in the presence of a strong bichromatic microwave field term, high fidelity gates can be implemented. This will likely enable experimental simplification, since one does not have to minimize the microwave magnetic field at the ions' positions.  Furthermore, the strength of the microwave magnetic field can be tuned to decouple the system from qubit frequency shifts without additional drive fields.

\begin{figure}[h]
\includegraphics[width=0.8\textwidth]{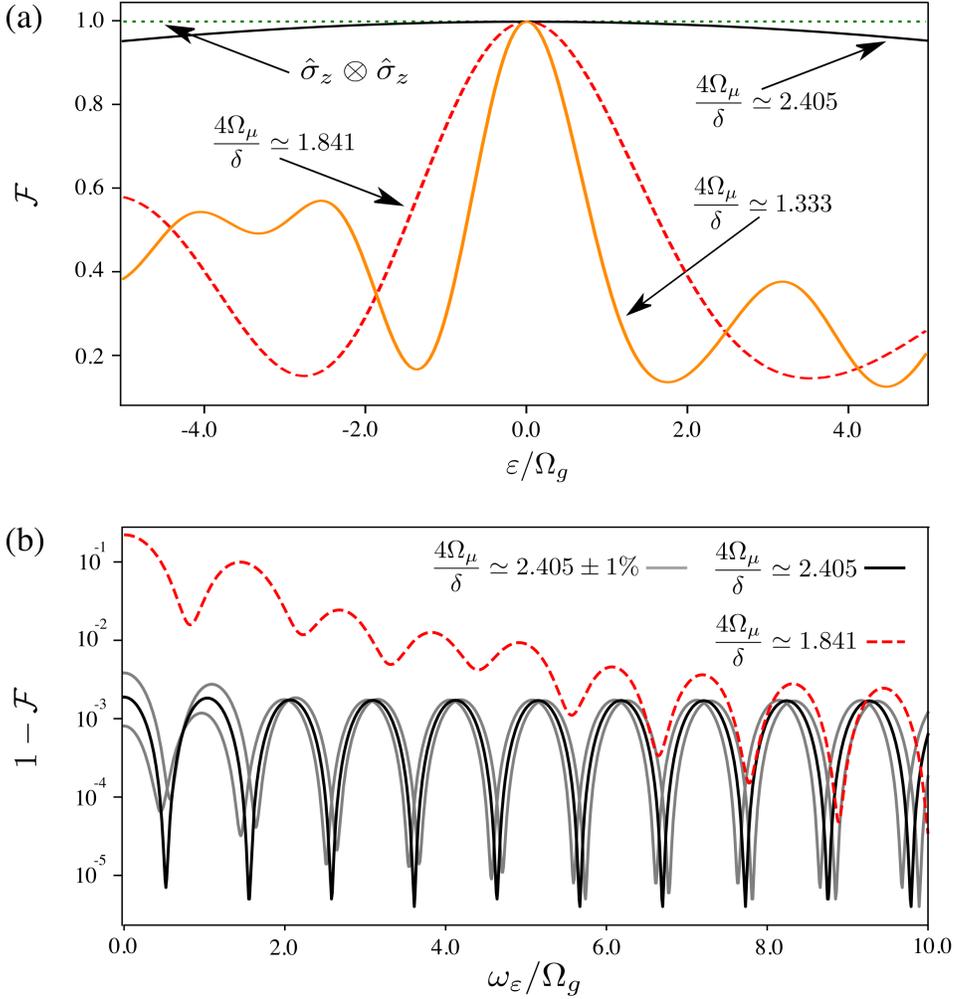}
\centering
\caption{(a) Fidelity $\mathcal{F}$ of the gate operation creating the maximally entangled Bell state of Eq.~(\ref{eq:bell}) versus static qubit frequency shift $\varepsilon$ normalized to gradient strength $\Omega_g$. Data in both panels are calculated by numerical integration of the full ion frame Hamiltonian given by Eq.~(\ref{eq:dynamic_ham}). Here, $\hat{S}_{i} = \hat{S}_{x}$, $\hat{S}_{j}=\hat{S}_{z}$, $\Omega_{g}/2\pi = 1$~kHz, and $\delta/2\pi = 1.5$~MHz (chosen to be experimentally realistic), with varying values of $\Omega_{\mu}$.  Fidelities are plotted for the intrinsically dynamically decoupled ${\hat{\sigma}_{\phi}\otimes\hat{\sigma}_{\phi}}$ gate ($4\Omega_{\mu}/\delta \simeq 2.405$, black solid line), the fastest ${\hat{\sigma}_{\phi}\otimes\hat{\sigma}_{\phi}}$ gate ($4\Omega_{\mu}/\delta \simeq 1.841$, red dashed line), the ${\hat{\sigma}_{\phi}\otimes\hat{\sigma}_{\phi}}$ gate shown in Fig.~\ref{fig:pulse_ms} ($4\Omega_{\mu}/\delta \simeq 1.333$, orange solid line), as well as the fastest ${\hat{\sigma}_{z}\otimes\hat{\sigma}_{z}}$ gate with a spin-echo ($4\Omega_{\mu}/\delta \simeq 3.054$, green dotted line) described in Sec.~\ref{sec:zz}. (b) Infidelity $1 - \mathcal{F}$ of the ${\hat{\sigma}_{\phi}\otimes\hat{\sigma}_{\phi}}$ gate versus the frequency $\omega_\varepsilon$ at which the qubit shift $\varepsilon$ oscillates, for the intrinsically dynamically decoupled gate (solid black) and the fastest ${\hat{\sigma}_{\phi}\otimes\hat{\sigma}_{\phi}}$ gate (red dashed), for a particular value of $\varepsilon_0 = \Omega_{g}$. This value of $\varepsilon_0$ represents a significantly larger qubit shift than is typically seen experimentally, where $|\varepsilon_0| \ll \Omega_{g}$ \cite{harty_2016}.  The grey lines show the effect of $\pm1\,\%$ relative changes in the ratio $4\Omega_{\mu}/\delta$ for the intrinsically dynamically decoupled gate.}
\label{fig:zeeman_shifts}
\end{figure}

The effect of this intrinsic dynamical decoupling on the gate is shown in Fig.~\ref{fig:zeeman_shifts}(a). Here, we plot $\mathcal{F}$ versus the normalized qubit frequency shift $\varepsilon/\Omega_g$ (for a static $\varepsilon$) assuming the parameters listed above, except that we now vary $\Omega_{\mu}$ to change the arguments of the Bessel functions. Figure~\ref{fig:zeeman_shifts}(a) shows that for most values of $\Omega_{\mu}$, $\mathcal{F}$ is highly sensitive to qubit frequency fluctuations. We plot $\mathcal{F}$ for the gate described above ($4\Omega_{\mu}/\delta \simeq 1.333$), and for a gate where $\Omega_{\mu}$ is increased to maximize the gate speed ($4\Omega_{\mu}/\delta \simeq 1.841$). For these two plots, we find that when $|\varepsilon|/\Omega_g\gtrsim 1$, the value of  $\mathcal{F}$ for the gate drops to $\sim 0.5$. However, when we further increase $\Omega_{\mu}$ such that $4\Omega_{\mu}/\delta \simeq 2.405$, i.e. the first root of the $J_{0}$ Bessel function, $\mathcal{F}$ becomes significantly less sensitive to $\varepsilon$, giving $\mathcal{F}\geq 0.95$ for $|\varepsilon|/\Omega_g \leq 5$. 

We can also take the $\varepsilon$ to be time-varying, of the form $\varepsilon = \varepsilon_0\cos(\omega_\varepsilon t)$.  Figure~\ref{fig:zeeman_shifts}(b) shows the dependence of the infidelity $1-\mathcal{F}$ on $\omega_\varepsilon$, assuming $\varepsilon_0 = \Omega_{g}$.  Infidelities are plotted for $4\Omega_{\mu}/\delta \simeq 2.405$ (intrinsic dynamical decoupling) and $4\Omega_{\mu}/\delta \simeq 1.841$ (fastest ${\hat{\sigma}_{\phi}\otimes\hat{\sigma}_{\phi}}$ gate). Figure~\ref{fig:zeeman_shifts}(b) shows that intrinsic dynamical decoupling protects against qubit energy shifts at frequencies up to $\approx 10\Omega_g$. This figure also shows the sensitivity of intrinsic dynamical decoupling to small fluctuations in $\Omega_{\mu}/\delta$; the grey lines show the infidelity when the ratio $4\Omega_{\mu}/\delta$ is shifted by $1\,\%$ above and below the intrinsic dynamical decoupling point respectively. By performing simulations for various values of $\varepsilon_0$, we determine that the infidelity scales as $(\varepsilon_0/\Omega_g)^2$ for $\varepsilon_0 \leq \Omega_g$.  

Unlike the dynamically decoupled ${\hat{\sigma}_{\phi}\otimes\hat{\sigma}_{\phi}}$ gate demonstrated in Ref.~\cite{harty_2016}, no additional field is required, and the microwave field term generating the dynamical decoupling \textit{does not} have to commute with the gradient term in the Hamiltonian. In fact, as will be discussed in the next section, the infinite series of resonances resulting from the microwave field term not commuting with the gradient provides the opportunity for a novel type of ${\hat{\sigma}_{z}\otimes \hat{\sigma}_{z}}$ microwave gate, where all frequencies are detuned from the ions' motional modes. 

\subsection{${\hat{\sigma}_{z}\otimes \hat{\sigma}_{z}}$ Gate}\label{sec:zz}

Dynamical decoupling can be beneficial for high-fidelity ${\hat{\sigma}_{\phi}\otimes\hat{\sigma}_{\phi}}$ gates, because the terms in the Hamiltonian that represent qubit frequency shifts do not commute with the gate. However, qubit frequency shifts commute with a ${\hat{\sigma}_{z}\otimes\hat{\sigma}_{z}}$ gate. Because of this, a simple spin-echo sequence completely cancels the effect of static qubit frequency shifts. Unfortunately, until now, the only proposed technique for performing microwave-based ${\hat{\sigma}_{z}\otimes\hat{\sigma}_{z}}$ gates requires generating oscillating gradients near the ions' motional frequencies, where experimental imperfections can give rise to electric fields that excite the ions' motion and reduce fidelity. This makes the gate difficult to perform in practice \cite{ospelkaus_2008}. 
\begin{figure}[h]
\includegraphics[width=0.8\textwidth]{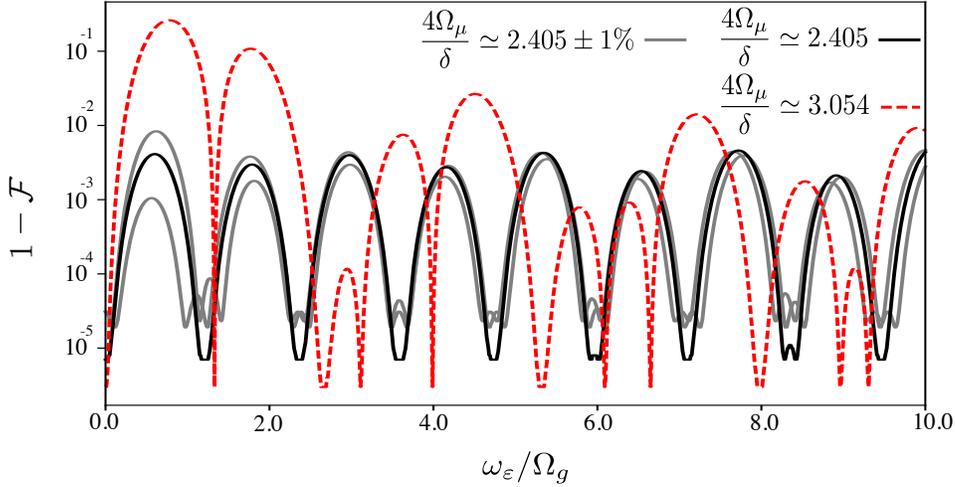}
\centering
\caption{Infidelity $1 - \mathcal{F}$ of the ${\hat{\sigma}_{z}\otimes\hat{\sigma}_{z}}$ gate versus the frequency $\omega_\varepsilon$ at which the qubit shift $\varepsilon$ oscillates, for the intrinsically dynamically decoupled gate (solid black) and the fastest ${\hat{\sigma}_{z}\otimes\hat{\sigma}_{z}}$ gate (red dashed), for a particular value of $\varepsilon_0 = \Omega_{g}$. This value of $\varepsilon_{0}$ represents a significantly larger qubit shift than is typically seen experimentally, where $|\varepsilon_0| \ll \Omega_{g}$ \cite{harty_2016}.  The grey lines show the effect of $\pm1\,\%$ relative changes in the ratio $4\Omega_{\mu}/\delta$ for the intrinsically dynamically decoupled gate.  Data are calculated by numerical integration of the full ion frame Hamiltonian given by Eq.~(\ref{eq:dynamic_ham}).}
\label{fig:zz_gates}
\end{figure}
When considering a gate in the bichromatic interaction picture as shown in Eq.~(\ref{eq:dynamic_jacobi}), a ${\hat{\sigma}_{z}\otimes\hat{\sigma}_{z}}$ gate is obtained when $2n\delta \sim |\omega_{r}-\omega_{g}|$. The $n=0$ case corresponds to a ${\hat{\sigma}_{z}\otimes\hat{\sigma}_{z}}$ gate with $\omega_g\simeq\omega_r$, as described above.  If instead we choose $n=1$, Eq.~(\ref{eq:dynamic_jacobi}) becomes:
\begin{equation}
\hat{H}_{I}(t) \simeq \hbar\Omega_{g}J_{2}\Big(\frac{4\Omega_{\mu}}{\delta }\Big)\hat{S}_{z}\Big\{\hat{a}e^{i\Delta t} + \hat{a}^{\dagger}e^{-i\Delta t} \Big\},
\end{equation}
where $\Delta \equiv 2\delta - (\omega_{r}-\omega_{g})$. We have now created a ${\hat{\sigma}_{z}\otimes \hat{\sigma}_{z}}$ gate where both $\omega_{g}$ and $\delta$ can be far-detuned from $\omega_{r}$, which relaxes the constraints on residual electric fields\footnote[1]{These electric fields drive the ion motion off-resonantly at $\omega_g$.  For a given electric field strength, the resulting motional amplitude scales as $1/(\omega_r^2-\omega_g^2)$.}.  Performing this gate with a spin-echo pulse after the first of two loops in phase space completely cancels the effect of the static shifts. This complete insensitivity to \textit{static} qubit shifts is demonstrated in Fig.~\ref{fig:zeeman_shifts}(a), while Fig.~\ref{fig:zz_gates} shows this gate's sensitivity to time-varying qubit shifts. While this ${\hat{\sigma}_{z}\otimes\hat{\sigma}_{z}}$ gate is less sensitive to static ($\omega_\varepsilon = 0$) noise, it remains sensitive to noise with larger values of $\omega_\varepsilon$. We note that intrinsic dynamical decoupling can also be applied to this gate, in addition to the spin echo.  In Fig.~\ref{fig:zz_gates}, we show simulated gate fidelities assuming $\varepsilon_0=\Omega_{g}$ at the intrinsically dynamically decoupled point (solid black) as well as for $4\Omega_{\mu}/\delta \simeq 3.054$ (dashed red), which gives the maximum relative gate speed $J_{2}\Big(3.054\Big) \simeq 0.49$; all other parameters are the same as described in Sec.~\ref{sec:phi_gate}. From other simulations, we determine that the infidelity scales as $(\varepsilon_{0}/\Omega_{g})^2$ for $\varepsilon_{0} \leq \Omega_g$.  

\section{Conclusion}\label{sec:conclusions}
This work analyzes microwave entangling gates in the bichromatic interaction picture, as opposed to the ion frame. This change of perspective offers key insights on how to implement experimental simplifications. If the microwave field does not commute with the gradient term in the Hamiltonian, an infinite series of resonances emerges in the bichromatic interaction picture. Individual resonances, selected by changing the microwave frequency, enable either ${\hat{\sigma}_{\phi}\otimes\hat{\sigma}_{\phi}}$ or ${\hat{\sigma}_{z}\otimes\hat{\sigma}_{z}}$ gates with all oscillating field frequencies far-detuned from the motional modes of the system.  In addition, the bichromatic microwave field amplitude can be tuned to provide intrinsic dynamical decoupling from qubit frequency fluctuations without additional fields.

\section*{Acknowledgements}
We acknowledge helpful discussions with D. Lucas, and thank S. Erickson,  A. Collopy, and Y. Rosen for careful reading of the manuscript.  R.S., S.C.B., and D.T.C.A. are Associates in the Professional Research Experience Program (PREP) operated jointly by NIST and the University of Colorado Boulder under award 70NANB18H006 from the U.S. Department of Commerce, National Institute of Standards and Technology.  This work was supported by ARO, ONR, and the NIST Quantum Information Program.  This paper is a partial contribution of NIST and is not subject to US copyright. Part of this work was performed under the auspices of the U.S. Department of Energy by Lawrence Livermore National Laboratory under Contract DE-AC52-07NA27344. LLNL-JRNL-759200

\section*{Appendix}
Here, we quantify the fidelity oscillations in Fig.~\ref{fig:pulse_ms}. We are interested in timescales that are fast compared to the gate speed, and therefore we neglect the gradient term in Eq.~(\ref{eq:dynamic_ham}). This gives:
\begin{equation}
\hat{H}(t) \simeq 2\hbar\Omega_{\mu}\cos(\delta t)\hat{S}_{x},
\end{equation}
making $\ket{\psi(t)} = \hat{U}(t)\ket{\psi(0)}$, where $\ket{\psi(0)} = \ket{Bell}$.  We note that this choice of $t=0$ (made for increased clarity in this derivation) is distinct from the main text, where $t=0$ represents the start time of the gate.  The state evolution is governed by
\begin{eqnarray}
\hat{U}(t) &=& \nonumber \exp\Big\{-\frac{i}{\hbar}\int^{t}_{0}dt^{\prime}H(t^{\prime}) \Big\} \\ 
&=& \exp\Big\{-\frac{2i\Omega_{\mu}}{\delta}\sin(\delta t)\hat{S}_{x} \Big\} \nonumber \\
&=& \Big\{\hat{I}\cos\Big(\frac{2\Omega_{\mu}}{\delta}\sin(\delta t)\Big) - i\hat{\sigma}_{x,1}\sin\Big(\frac{2\Omega_{\mu}}{\delta}\sin(\delta t) \Big) \Big\}\Big\{\hat{I}\cos\Big(\frac{2\Omega_{\mu}}{\delta}\sin(\delta t)\Big) \nonumber \\ && - i\hat{\sigma}_{x,2}\sin\Big(\frac{2\Omega_{\mu}}{\delta}\sin(\delta t) \Big) \Big\}.
\end{eqnarray}
Figure~\ref{fig:pulse_ms} shows the fidelity $\mathcal{F}$ of the maximally entangled Bell state $\ket{Bell} \equiv 2^{-1/2}(\ket{\downarrow\downarrow} - i\ket{\uparrow\uparrow})$. Here the value of $\mathcal{F}$ at time $t$ is given by:
\begin{eqnarray}
\mathcal{F}(t) = |\bra{Bell}\hat{U}(t)\ket{Bell}|^2.
\end{eqnarray}
Keeping in mind that $\bra{Bell}\hat{\sigma}_{x,i}\ket{Bell} = \bra{Bell}\hat{\sigma}_{x,1}\hat{\sigma}_{x,2}\ket{Bell} = 0$, we obtain
\begin{eqnarray}
\mathcal{F}(t) &=& |\bra{Bell}\Big\{\hat{I}\cos^{2}\Big(\frac{2\Omega_{\mu}}{\delta}\sin(\delta t)\Big)\Big\}\ket{Bell}|^{2} \nonumber \\
&=& \cos^{4}\Big(\frac{2\Omega_{\mu}}{\delta}\sin(\delta t) \Big),
\end{eqnarray}
which corresponds to the frequency and magnitude of the oscillations in Fig.~\ref{fig:pulse_ms}.

\section*{References}
\bibliography{bichromatic}
\bibliographystyle{iopart-num}
\end{document}